\documentclass{ptephy_v1} %%%%%% to generate preprint number with ptep logo

\preprintnumber{lhep-2002} %%% %%% Insert preprint number here
\usepackage{lineno}

\begin{document}

\title{Slow control and monitoring system at the JSNS$^{2}$}
%\linenumbers
%%%% To generate auto affiliation numbers please use \author{}\affil{} command
\author[9]{J. S. Park}
\author[15]{S.~Ajimura} 
\author[14]{M.~Botran}
\author[4]{J.~H.~Choi}
\author[3]{J,~W.~Choi}
\author[18]{M.~K.~Cheoun}
\author[21]{T.~Dodo}
\author[21]{H.~Furuta}
\author[12]{J.~Goh}
\author[7]{M.~Harada} 
\author[7]{S.~Hasegawa} 
\author[21]{Y.~Hino}
\author[15]{T.~Hiraiwa} 
\author[16]{H.~I.~Jang}
\author[6]{J.~S.~Jang}
\author[3]{M.~C.~Jang}
\author[19]{H.~Jeon}
\author[19]{S.~Jeon}
\author[3]{K.~K.~Joo}
\author[14]{J.~R.~Jordan} 
\author[19]{D.~E~Jung} 
\author[17]{S.~K.~Kang}
\author[7]{Y.~Kasugai} 
\author[10]{T.~Kawasaki} 
\author[8]{E.~J.~Kim}
\author[3]{J.~Y.~Kim}
\author[19]{S.~B.~Kim}
\author[13]{W.~Kim}
\author[10]{T.~Konno}
\author[9]{D.~H.~Lee}
\author[12]{S.~Lee}
\author[3]{I.~T.~Lim}
\author[14]{E.~Marzec} 
\author[9]{T.~Maruyama}
\author[7]{S.~Meigo} 
\author[9]{S.~Monjushiro}
\author[3]{D.~H.~Moon}
\author[15]{T.~Nakano} 
\author[11]{M.~Niiyama}
\author[9]{K.~Nishikawa{\footnote{Deceased}}}
\author[15]{M.~Nomachi} 
\author[4]{M.~Y.~Pac{\footnote{Corresponding author, email:pac@dsu.ac.kr}}}
\author[20]{S.~J.~M.~Peeters}
\author[5]{H.~Ray}
\author[19]{G.~Roellinghoff}
\author[19]{C.~Rott} 
\author[7]{K.~Sakai} 
\author[7]{S.~Sakamoto}
\author[15]{T.~Shima} 
\author[3]{C.~D.~Shin}
\author[14]{J.~Spitz} 
\author[1]{I.~Stancu}
\author[15]{Y.~Sugaya}
\author[21]{F.~Suekane}
\author[7]{K.~Suzuya}
\author[9]{M.~Taira}
\author[21]{R.~Ujiie}
\author[2]{M.~Yeh}
\author[16]{I.~S.~Yeo}
\author[12]{C.~Yoo}
\author[19]{I.~Yu} 
\author[3]{A.~Zohaib}

\affil[1]{\small{University of Alabama, Tuscaloosa, AL, 35487, USA}}
\affil[2]{Brookhaven National Laboratory, Upton, NY, 11973-5000, USA}
\affil[3]{Department of Physics, Chonnam National University, Gwangju, 61186, KOREA}
\affil[4]{Laboratory for High Energy Physics, Dongshin University, Chonnam 58245, KOREA}
\affil[5]{University of Florida, Gainesville, FL, 32611, USA}
\affil[6]{Gwangju Institute of Science and Technology, Gwangju, 61005, KOREA}
\affil[7]{J-PARC Center, JAEA, Tokai, Ibaraki 319-1195, JAPAN}
\affil[8]{Division of Science Education, Jeonbuk National University, Jeonju, 54896, KOREA}
\affil[9]{High Energy Accelerator Research Organization (KEK), Tsukuba, Ibaraki 305-0801, JAPAN}
\affil[10]{Department of Physics, Kitasato University, Sagamihara, Kanagawa 252-0373, JAPAN}
\affil[11]{Department of Physics, Kyoto Sangyo University, Kyoto 603-8555, JAPAN}
\affil[12]{Department of Physics, Kyung Hee University, Seoul 02447, Korea}
\affil[13]{Department of Physics, Kyungpook National University, Daegu 41566, KOREA}
\affil[14]{University of Michigan, Ann Arbor, MI, 48109, USA}
\affil[15]{Research Center for Nuclear Physics, Osaka University, Osaka 565-0871, JAPAN}
\affil[16]{Department of Fire Safety, Seoyeong University, Gwangju 61268, KOREA}
\affil[17]{School of Liberal Arts, Seoul National University of Science and Technology, Seoul, 139-743, KOREA}
\affil[18]{Department of Physics, Soongsil University, Seoul 06978, KOREA}
\affil[19]{Department of Physics, Sungkyunkwan University, Suwon 16419, KOREA}
\affil[20]{Department of Physics and Astronomy, University of Sussex, BN1 9QH, Brighton, UK}
\affil[21]{Research Center for Neutrino Science, Tohoku University, Sendai, Miyagi 980-8577, JAPAN}

\begin{abstract}
The JSNS$^2$ experiment aims to search for sterile neutrino oscillations using a neutrino beam from muon decays at rest. The JSNS$^2$ detector contains 17 tons of 0.1\% gadolinium (Gd) loaded liquid scintillator (LS) as a neutrino target. Detector construction was completed in the spring of 2020. A slow control and monitoring system (SCMS) was implemented for reliable control and quick monitoring of the detector operational status and environmental conditions. It issues an alarm if any of the monitored parameters exceed a preset acceptable range. The SCMS monitors the high voltage (HV) of the photomultiplier tubes (PMTs), the LS level in the detector, possible LS overflow and leakage, the temperature and air pressure in the detector, the humidity of the experimental hall, and the LS flow rate during filling and extraction.  An initial 10 days of data-taking with a neutrino beam was done following a successful commissioning of the detector and SCMS in June 2020. In this paper, we present a description of the assembly and installation of the SCMS and its performance.

\end{abstract}

\subjectindex{Detector control systems (detector and experiment monitoring and slow-control systems, architecture, hardware, algorithms, databases), Neutrino detectors}

\maketitle

\section{Introduction}
The Sterile Neutrino Search at J-PARC Spallation Neutron Source (JSNS$^2$) experiment aims to search for sterile neutrino oscillation with mass-squared difference $\Delta$m$^{2}$ near 1\,eV$^{2}$ at J-PARC~\cite{cite:JSNS2_proposal}. The experiment uses a pulsed neutrino beam created from muon decays at rest, where the muons are produced from collisions of a 3 GeV proton beam on a mercury target. Based on this source a sensitive search for muon antineutrino oscillations to electron antineutrino is possible. The JSNS$^2$ detector is located on the third floor of the material and life science facility (MLF), 24 m away from the target of neutrino source \cite{cite:JSNS2_TDR}. 

\begin{figure}
\begin{center}
\includegraphics[scale=0.4]{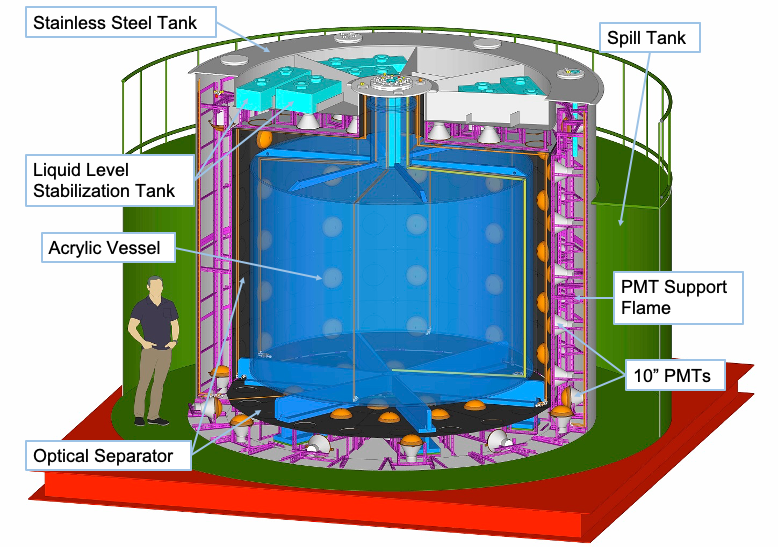}
\end{center}
\caption{\setlength{\baselineskip}{4mm}Configuration of the JSNS$^2$ detector.}
\label{fig:position}
\end{figure}

The detector consists of 17 tons of Gd loaded LS (Gd-LS), as a neutrino target, in an acrylic vessel, and 31 tons of unloaded LS as a gamma-catcher and a veto, in a stainless steel container surrounding the target. The gamma-catcher and veto are optically separated. Scintillation light produced in the target and gamma-catcher regions is detected by an array of 96 10-inch PMTs  \cite{cite:JSNS2_PMT}. Light produced in the veto region is detected by an array of 24 10-inch PMTs. Figure~\ref{fig:position} shows an overview of the JSNS$^2$ detector.

\begin{figure}[h]
\begin{center}
\includegraphics[scale=0.5]{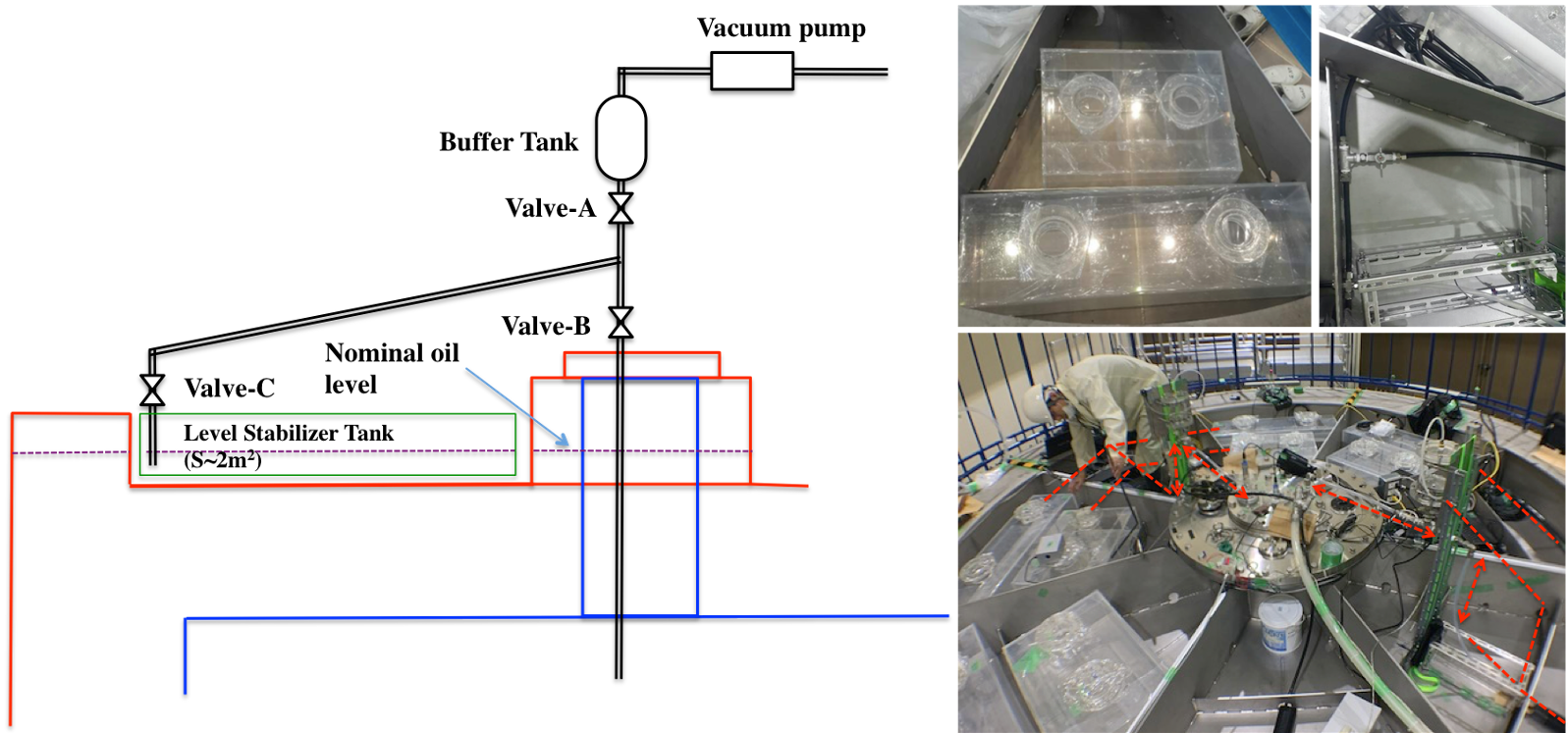}
\end{center}
\caption{\setlength{\baselineskip}{4mm}Conceptual design of the stabilizer container (left) and photos taken after deployment on the detector top lid (right). In the left design, the red, blue and green lines indicate the stainless steel tank, the target acrylic vessel, and the stabilizer container, respectively. The right pictures show a set of two acrylic containers (top left), connecting pipes between container sets (top right), and four sets of containers are installed on the detector top lid (bottom). The red dashed arrows represent the pipes connecting the stabilizer containers to the acrylic vessel inside the detector via the inverse siphon system.}
\label{fig:stabilizer}
\end{figure}

High voltage is supplied to each of the 120 PMTs individually and can be monitored and adjusted to maintain their uniform and constant gains. Since the MLF uses a highly irradiated mercury target under the JSNS$^2$ detector, the safety requirement for the JSNS$^2$ detector is quite stringent. The flammable LS must be treated carefully according to the Fire Law in Japan. The LS level in the detector is monitored by ultrasonic sensors. Several stabilization containers are installed on the top of the detector for the purpose of preventing the Gd-LS from overflowing due to its thermal expansion as shown in Fig.~\ref{fig:stabilizer}. The Gd-LS levels in the stabilization containers are also monitored by ultrasonic sensors. The temperature and pressure in the detector and the humidity of the experimental hall are measured and monitored by installed sensors. The detector is surrounded by the spill tank and any LS spill is detected by ultrasonic sensors and web cameras. 

The third floor of the MLF is a maintenance area for the spallation neutron target and beam line equipment. The JSNS$^2$ detector must be relocated from the third floor during a maintenance period, typically from July to September. For detector relocation, the LS must be filled into, or extracted from, the detector. The LS flow rate and level in the detector are carefully monitored during filling and extraction.

After an overview of the slow control and monitoring system (SCMS) in section 2, we describe the HV control and monitoring in section 3. Monitoring of the LS level in the detector, possible LS overflow and leakage, and LS flow rate during filling and extraction are discussed in section 4. Section 5 details our system for monitoring the temperature and pressure in the detector and experimental hall. Visualization and display of the SCMS data are discussed in section 6. 

\section{Overview of SCMS}
\begin{figure}[h]
\begin{center}
\includegraphics[scale=0.4]{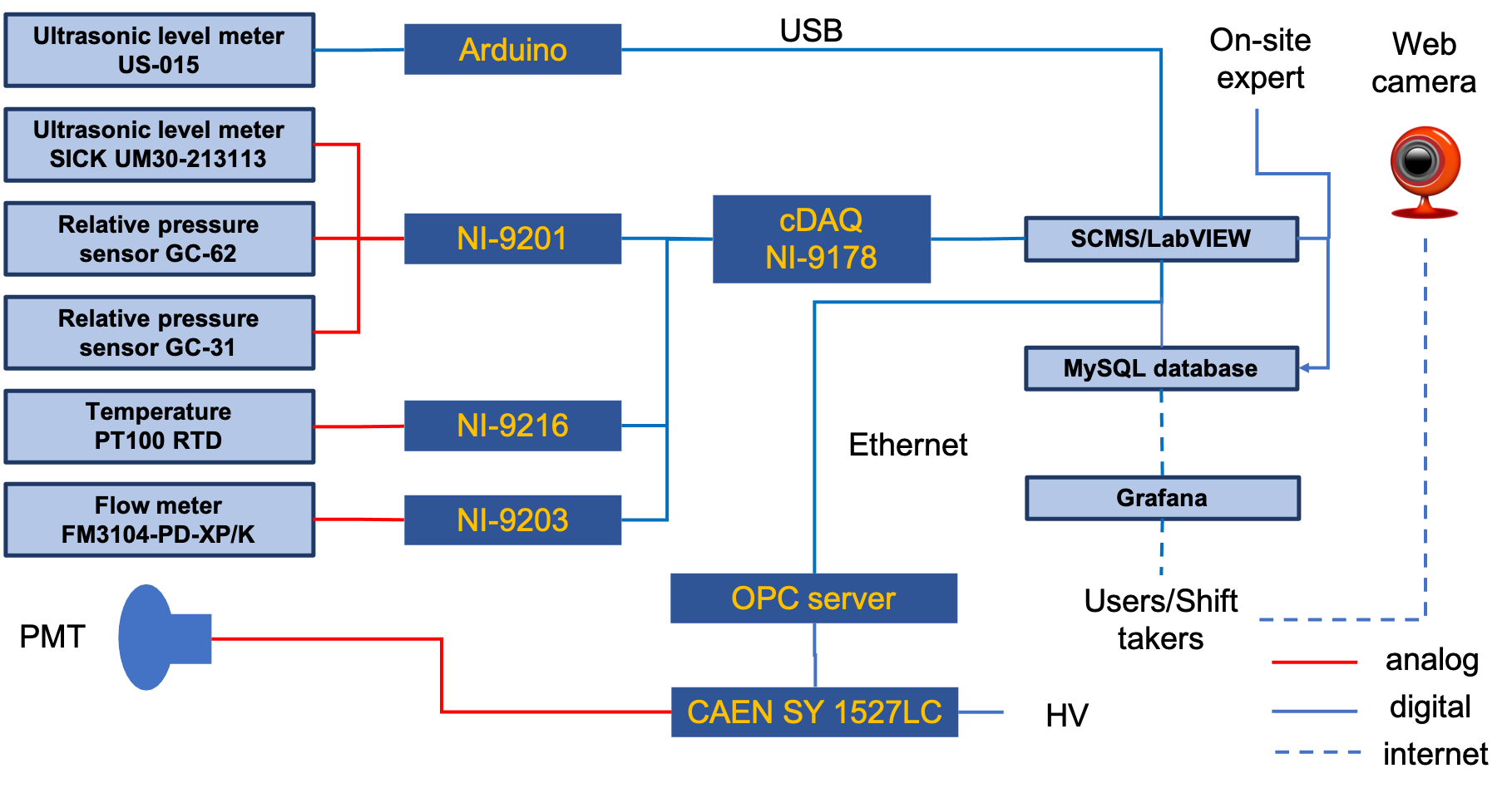}
\end{center}
\caption{\setlength{\baselineskip}{4mm}Conceptual diagram of the JSNS$^2$ slow control and monitoring system. The red and blue solid lines represent analog and digital signals, respectively. Except for Grafana, all components are connected via the local network. The camera is connected to the internet and used for monitoring the current status of the detector in real time.}
\label{fig:schematic}
\end{figure}

The SCMS provides reliable control and quick monitoring of detector operational status and environmental conditions. The SCMS can also issue alarms if any monitored values exceed a preset range, so that immediate maintenance can be done. As described in the previous section, it includes control and monitoring of HV supplied to PMTs, and monitoring of the LS level in the detector, possible LS overflow and leakage, the temperature and air pressure in the detector, the humidity of the experimental hall, and the LS flow rate during filling and extraction. The data acquired from the various sensors are delivered to a client program via a local network. The LabVIEW-based client program \cite{cite:LabVIEW} records the data into a MySQL \cite{cite:MySQL} database once every 30 seconds and displays the monitoring system's data for on-site experts. Outside the MLF, Grafana \cite{cite:Grafana}, receiving data from the client program, displays the detector's current status and its environment and generates alarms and control signals as needed. Figure~\ref{fig:schematic} shows a schematic drawing of the SCMS. Table 1 lists sensors for measuring various parameters and their readout modules.

\begin{table}[h]
\caption{\setlength{\baselineskip}{4mm}Monitored parameters and readout modules for used sensors.}
\begin{center}
\begin{tabular}{c | c | c}
\hline
Sensor (Model ID) & Measured quantity & Readout module  \\ \hline \hline
CAEN OPC server with SY1527LC & PMT high voltage & ethernet \\ \hline
Ultrasonic  & Liquid height & NI 9201\\ 
(SICK UM30-215113) & & \\ \hline
Ultrasonic  & Liquid height & NI 9201 \\
(SICK UM30-213113) & & \\ \hline
Ultrasonic & Liquid height inside & Arduino \\ 
 (HALJIA US-015)&the spill tank&\\ \hline
Ultrasonic  & Liquid height of  & Arduino \\ 
(HALJIA US-015)&the level stabilization&\\ \hline
Temperature (PT100 RTD)& LS temperature & NI 9216 \\ \hline
Relative pressure & Pressure between chimney  & NI 9201 \\ 
(GC-62)& and veto volume&\\ \hline
Relative pressure & Pressure between detector  & NI 9201 \\ 
(GC-31)&and laboratory&\\ \hline
Flow meter  & LS flow rate & NI 9203 \\ 
 (FM3104-PD-XP/K)&&\\ \hline
Ambient sensor  & Laboratory temperature,  & RS232 to USB\\ 
(TR73-U)&humidity and pressure& \\ \hline
Web camera & Liquid leakage inside  & N/A \\ 
& oil protection wall \\ \hline
\end{tabular}
\end{center}

\label{tab:hall_sensors}
\end{table}

The data acquired by several sensors are collected by readout modules and delivered to the LabVIEW client via USB cables. A National Instruments (NI) module 9216~\cite{cite:ni9216} is used to read data out from the resistance temperature detectors (RTDs). An NI 9201~\cite{cite:ni9201} reads analog voltage values from a number of sensors. An NI 9203 reads an analog current value from a flow meter. An NI cDAQ-9178~\cite{cite:ni9178} crate houses these three NI modules and is connected to an SCMS PC via a USB cable. LabVIEW was selected as a readout program because it interfaces well with the NI readout modules. 

Several Arduino modules described in section 4.2 are used to read data from ultrasonics sensors for detecting a possible liquid spill and monitoring the LS level in the stabilization containers, installed on top of the detector as shown in Fig. 2. These Arduino systems communicate with the LabVIEW program on the SCM PC via USB.

\begin{figure}[h]
\begin{center}
\includegraphics[scale=0.4]{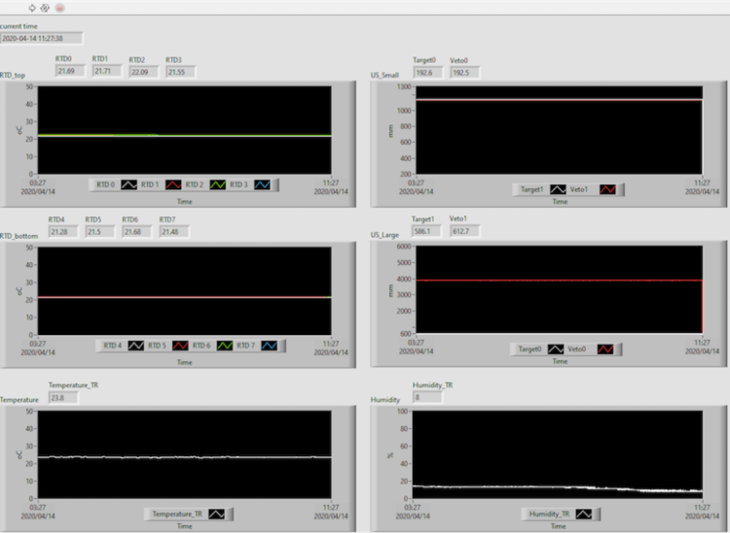}
\end{center}
\caption{\setlength{\baselineskip}{4mm}An example screenshot of the LabVIEW display.}
\label{fig:slow_monitor}
\end{figure}

The LabVIEW program queries each sensor’s measured value every 5 seconds and displays them over an 8-hour time span, and sends the data to MySQL database at the same time. The LabVIEW program can display information about the current status of the detector on the screen. It includes the liquid levels from four SICK ultrasonic sensors, which will be described in section 4.1, and six Arduino systems, the detector temperatures from eight RTDs (resistance temperature detectors), the detector pressure, and the humidity in the experimental hall. It also shows the liquid flow rate during filling and extraction. Figure~\ref{fig:slow_monitor} shows a typical part of screenshots of the LabVIEW display, which includes the RTD's detector temperature, the liquid levels from the ultrasonic sensors, and the experimental hall's temperature and humidity.  

\begin{figure}[h]
\begin{center}
\includegraphics[scale=0.4]{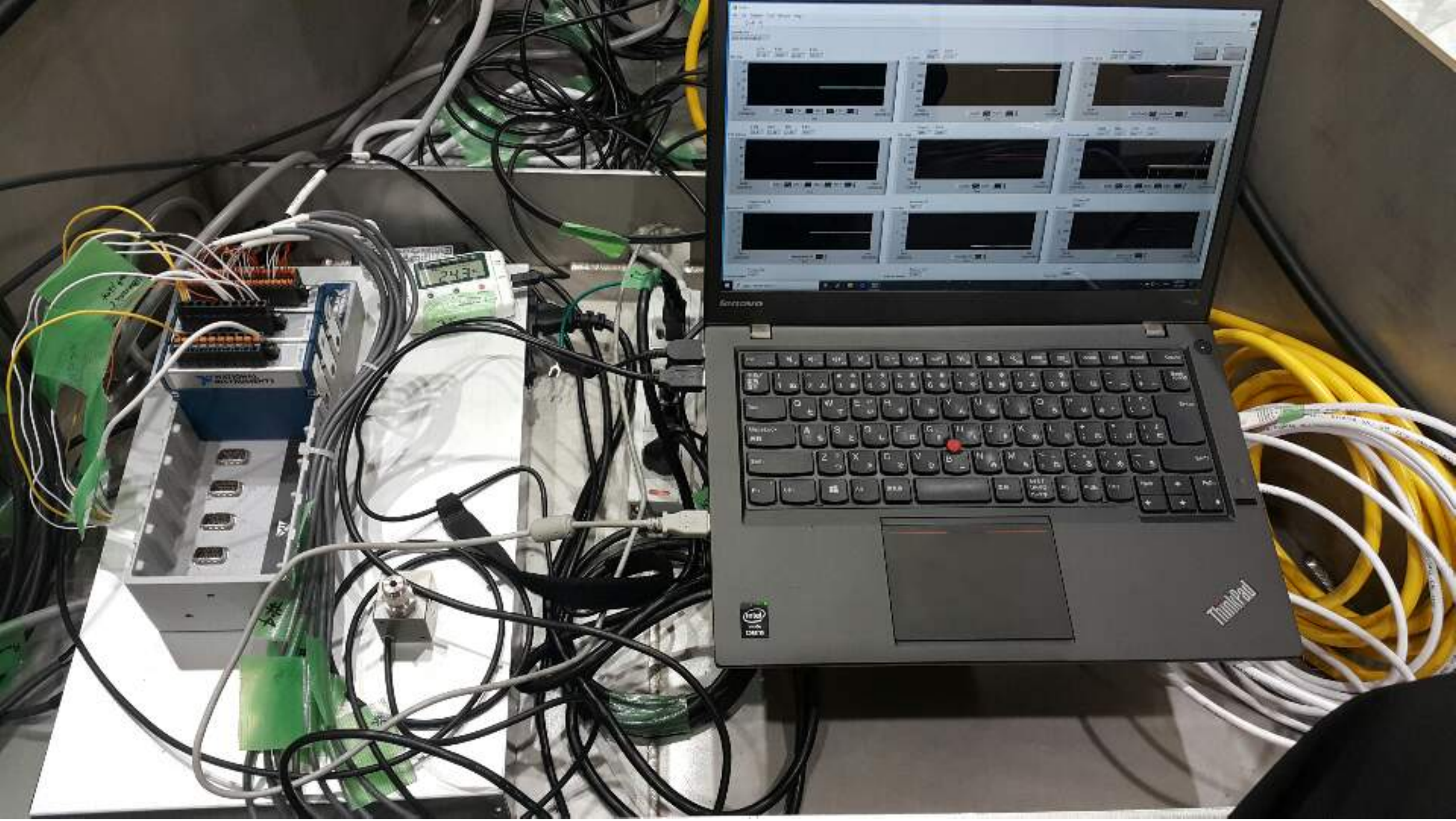}
\end{center}
\caption{\setlength{\baselineskip}{4mm} Analog modules (NI-9201, 9216, and 9203) installed in the NI cDAQ 9178 chassis (left) and SCMS system test PC (right).}
\label{fig:NI_modules}
\end{figure}

The LabVIEW program communicates with a MySQL database over a network connection to store the slow control and HV values in a single table every 30 seconds. The current HV value of each PMT is recorded on its own table. The temperature of each HV supplier module is stored in a dedicated table.

This system can be accessed through the network to display or record both current and historical data. The SCMS client applications allow users to manage and access the status of the experiment through a flexible graphical-user-interface based tool, Grafana. Figure~\ref{fig:NI_modules} shows the NI readout modules and the NI crate.

Table 2 shows each parameter's requirements and performance required for SCMS to perform the JSNS2 experiment successfully.

\begin{table}[h]
\caption{\setlength{\baselineskip}{4mm}Summary of the requirements for measuring SCMS parameters and the precision of the sensors.}
\begin{center}
\begin{tabular}{ l | p{5.5cm} | p{4.5cm}}
\hline
Parameter & Requirement & Performance  \\ \hline \hline
High voltage & Control PMT HV and issue an alarm for an abnormal channel.& $\pm ~0.5$ V $\pm~ 0.3\%$ (reading)

$\pm ~0.25$ V$~\pm~ 0.3\%$ (setting)\\ \hline
Liquid temperature  & Detect its abrupt change before a possible liquid overflow. & 0.15 $\pm$ 0.002T[$^\circ$C]\\ \hline
Liquid height &Ensure its difference between target and gamma-catcher less than 15 cm.& UM30-213113 : $\pm$1\%

UM30-215113 : $\pm$1\%\\ \hline
Detector air pressure&Ensure its difference between the detector inside and outside less than 20 kPa.&GC31 : $\pm$ 1.0\%

GC62 : $\pm$ 1.0\% \\ \hline
Liquid flow rate&Adjust the liquid filling and extraction speed to ensure the liquid level difference between target and gamma-catcher less than 15 cm.&\\ \hline
Oil leakage&Detect any oil leak as early as possible.&US-015 : $\pm$~0.1 cm ~$\pm$~1\%\\ \hline
Ambient parameters&Monitor environmental conditions possibly causing detector operation.&TR-73U:

Temperature : $\pm$~0,3~$^\circ$C

Humidity : $\pm$~5\% RH

Pressure : $\pm$~1.5hPa\\ \hline
\end{tabular}
\end{center}

\label{tab:requirement}
\end{table}

\section{High voltage control and monitoring}
The CAEN SY1527LC crate with six A1535 modules \cite{cite:CAEN_HV}, obtained from the Double Chooz experiment are reused to supply HV for 120 PMTs. A CAEN A1535 module can supply 24 different voltages for each channel up to 3.5 kV. The CAEN operational process control (OPC) server is used to control and communicate with the HV modules \cite{cite:CAEN_OPC}.  

\begin{figure}[h]
\begin{center}
\includegraphics[scale=0.5]{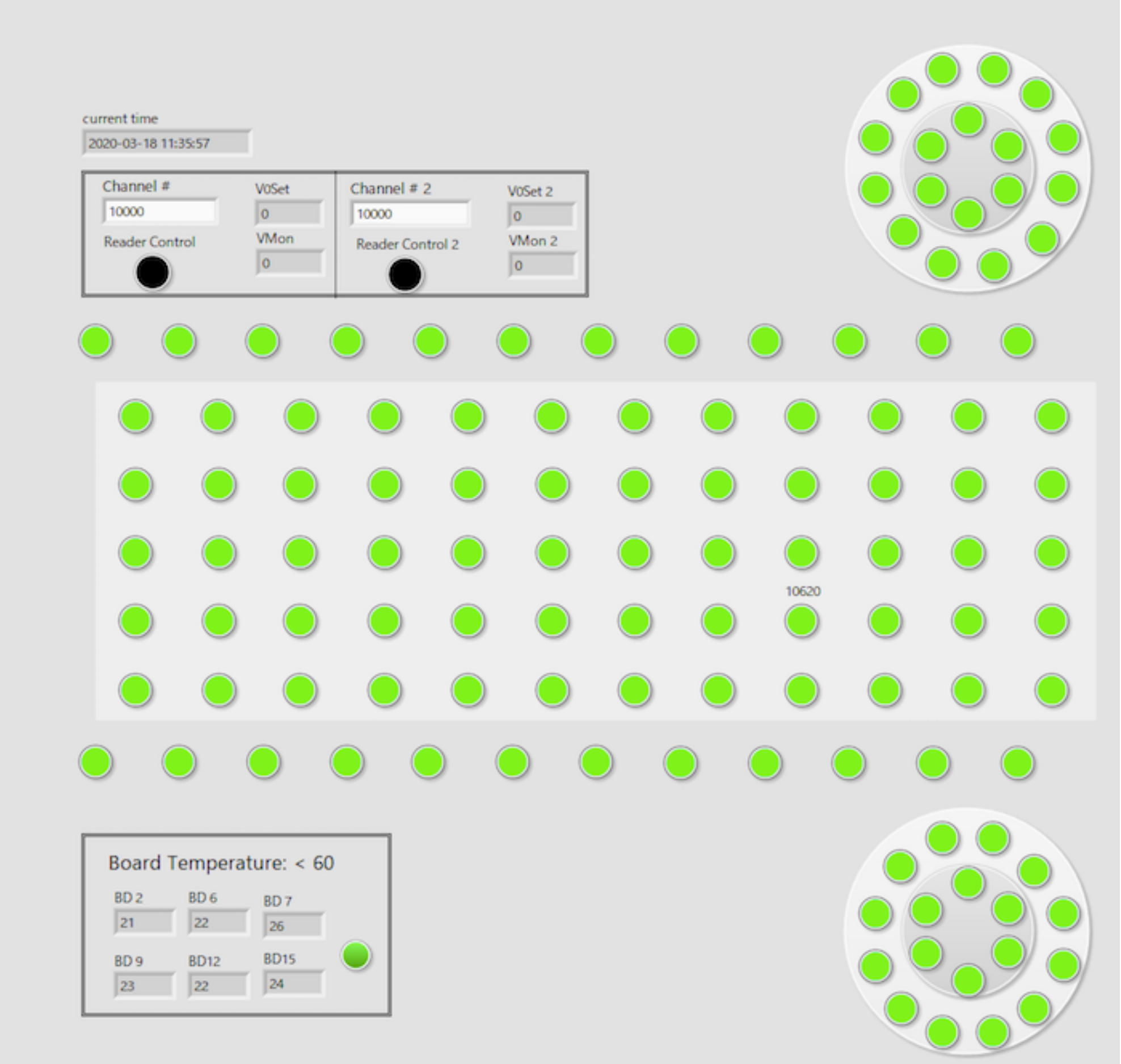}
\end{center}
\caption{\setlength{\baselineskip}{4mm}Screenshot of HVCM display. Each green circle indicates that the PMT voltage is within 10 V difference from its preset value. The temperature of each HV modules is shown in the bottom left.}
\label{fig:HV_LabVIEW}
\end{figure}

The CAEN SY1527LC crate is accessed via the OPC server from a dedicated, LabVIEW based, HV control and monitoring (HVCM) program. The HVCM program downloads a preset HV value for each channel, and stores the currently supplied HV of each channel and the temperature of each HV module. The HVCM program also displays the status of supplied PMT HV on a map of PMT locations. Figure~\ref{fig:HV_LabVIEW} shows a screenshot of the monitored HV status using the HVCM program. The color of each circle represents the HV status based on the difference between the preset and currently supplied values.

\section{Liquid scintillator monitor}
\begin{figure}
\begin{center}
\includegraphics[scale=0.6]{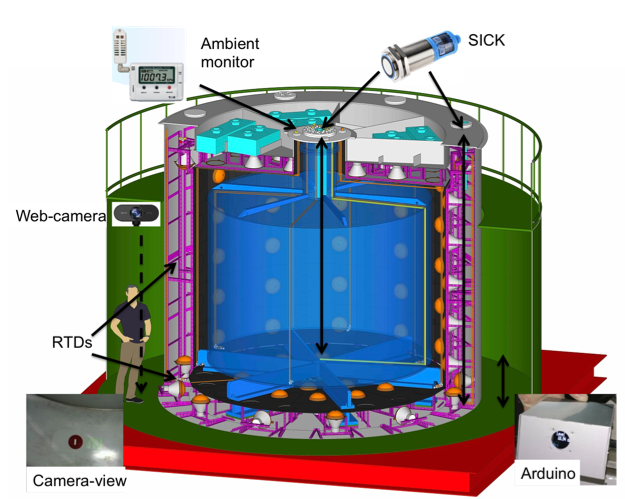}
\end{center}
\caption{\setlength{\baselineskip}{4mm}Locations of installed sensors.}
\label{fig:position_sensor}
\end{figure}

As described earlier, the Gd-LS and LS levels in the detector and stabilization containers are monitored by ultrasonic sensors. Possible liquid spill within the spill tank surrounding the detector is monitored by ultrasonic sensors and web cameras. The temperature of the LS is monitored by eight sensors installed inside the detector. The liquid flow rate and liquid levels in the detector are monitored during filling and extraction as well. Several kinds of ultrasonic sensors are used to measure the liquid levels according to their precisions and operational ranges. Figure~\ref{fig:position_sensor} shows installed sensor locations.  
\subsection{Liquid scintillator level monitor}
\begin{figure}[h]
\begin{center}
\includegraphics[scale=0.5]{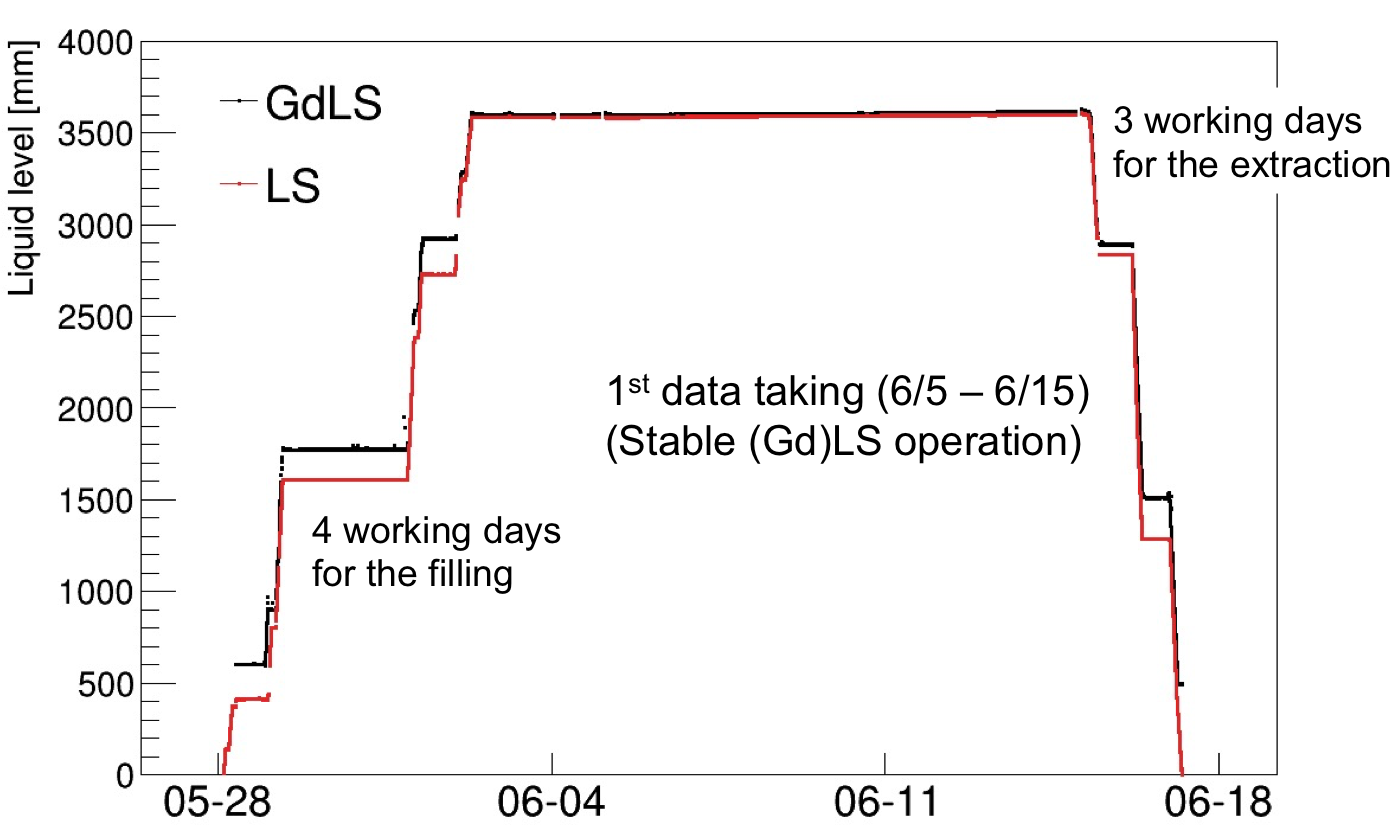}
\end{center}
\caption{\setlength{\baselineskip}{4mm}Variation of liquid level in the detector measured by ultrasonic sensors during liquid filling, data-taking, and liquid extraction.}
\label{fig:filling_extraction}
\end{figure}

SICK ultrasonic level meters \cite{cite:SICK} are used to monitor for the levels of Gd-LS and LS in the detector. A pair of two different sensors are installed on the detector chimney for the Gd-LS level monitoring and on the veto flange for the LS \cite{cite:SICK_um30_215113,cite:SICK_um30_213113}. 
The measurement of liquid level by ultrasonic sensors are interrupted by cables through the inlet and the PMT structure in the detector. An acrylic pipe with the sensors attached to it is used to avoid the interruption. The end of the acrylic pipe is carefully polished to avoid unnecessary reflection of ultrasonic waves. 
Each liquid level meter provides an analog voltage output from 0 to 10 V, proportional to the measured distance to the liquid surface; an NI 9201 module reads the analog voltage output. Each level meter also displays the measured distance on an LED screen. Figure~\ref{fig:filling_extraction} shows a distribution of measured liquid levels during filling, data-taking, and extraction modes.

\subsection{Liquid scintillator leak monitor}
There are four additional ultrasonic level sensors for monitoring the interior of the spill tank, and two additional ultrasonic level sensors for monitoring the liquid level inside the stabilization containers. The US-015 \cite{cite:US_015} ultrasonic sensor is used to detect even a relatively small amount of liquid in the spill tank. 
The US-015 is well suited for this because its operational range is from 20 mm to 4,000 mm and its resolution of roughly 1 mm. The sensors are installed about 10 cm from the floor, looking down as shown in Fig.~\ref{fig:position_sensor}. The sensors for the stabilization containers must be sensitive to the distances from 50 mm to 300 mm, and thus the US-015 sensors are also used for monitoring the liquid level inside the stabilization containers. 

\begin{figure}[h]
\begin{center}
\includegraphics[scale=0.5]{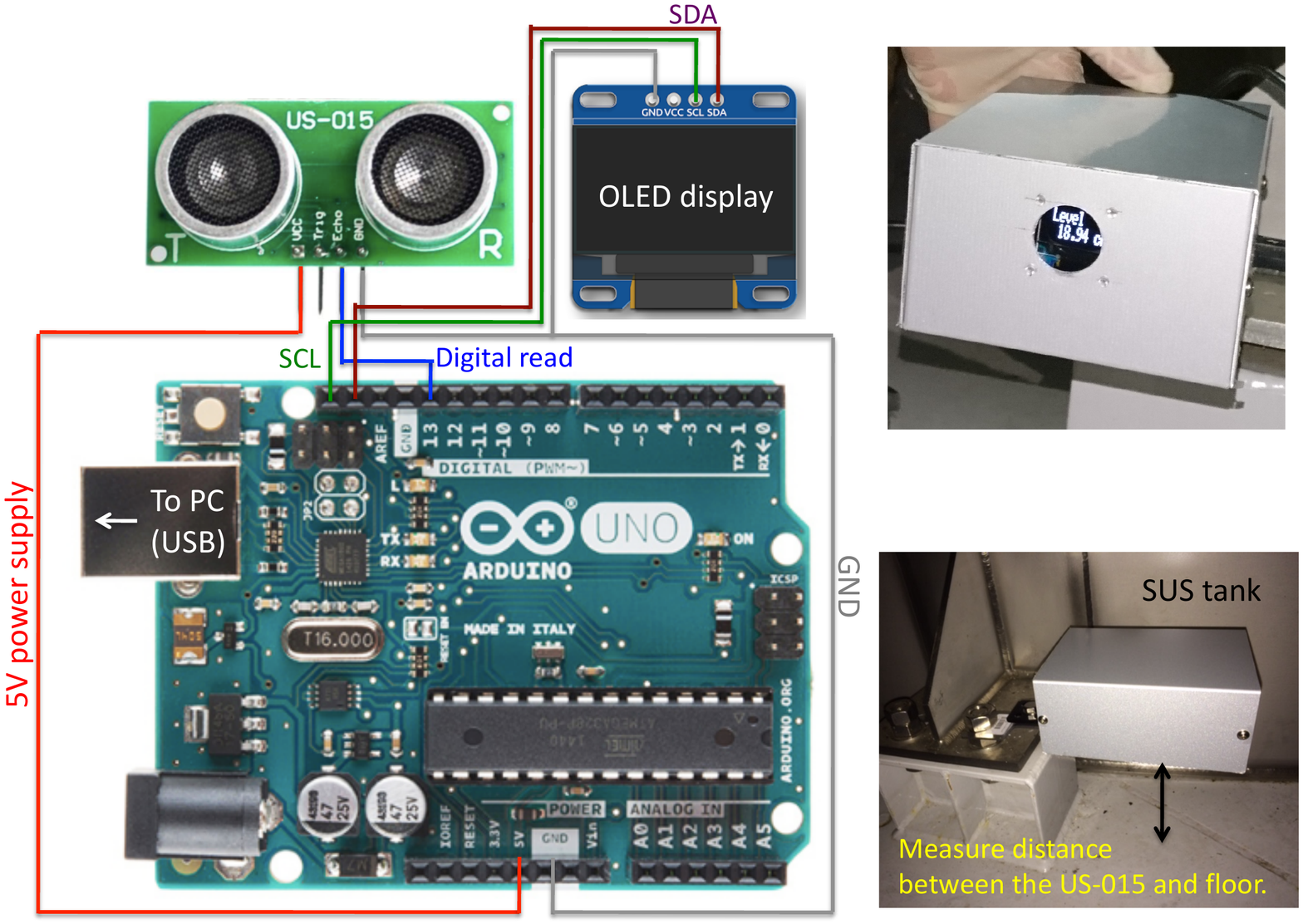}
\end{center}
\caption{\setlength{\baselineskip}{4mm}Liquid level monitoring device consisting of a US-015 ultrasonic sensor, an Arduino I2C UNO readout module, and an OLED display. A US-015 has a digital output related to the pulse height of information about the distance, and the SCMS reads digital values transmitted from Arduino via USB cable. The connection diagram among them is shown in the left photo. The assembled device in a housing is shown in the top right photo. The bottom right photo shows the device installed for the spill tank.}
\label{fig:Arduino}
\end{figure}

The US-015 sensors are read out by an Arduino I2C Uno Rev3 module \cite{cite:Arduino}, a micro-controller capable of analog-to-digital conversion. The Arduino module sends data to the SCM PC over a USB connection. The obtained data are displayed on a connected 0.96-inch organic LED that is produced by SUNHOKEY Electronics Co. Ltd. \cite{cite:OLED}. Figure~\ref{fig:Arduino} shows a liquid level monitoring device consisting of an ultrasonic sensor, an Arduino readout module, and an OLED display for a stabilization container \cite{cite:solomon}. It also shows a liquid level monitoring device for the spill tank.

\subsection{Monitoring for liquid scintillator filling and extraction}
As described earlier, the JSNS$^2$ detector is installed on the third floor of the MLF building. The area is reserved for regular maintenance of the mercury target and a beam line equipment. During the maintenance period, the detector must be removed from the MLF and stored elsewhere. The (Gd-)LS must be filled or extracted before installation or relocation, respectively. Both Gd-LS and LS levels inside the detector must be maintained as evenly as possible to minimize the stress on the acrylic vessel.

An FM3104-PD-XP/K \cite{cite:Flow_meter} flow meter is used to measure the liquid flow rate into or out of the detector during filling and extraction, respectively. A frequency inverter for the pump is used to modulate the flow rate of liquid. The flow meter displays the measured flow rate on an LED display and provide an analog current output from 4 to 20 mA, proportional to the flow rate. An NI 9203 module reads and digitizes the analog current output. 

\section{Monitors for detector temperature, pressure and environmental condition}
\subsection{Detector temperature monitor}
Thermal expansion of the liquid scintillator could result in overflow from the detector. Eight stabilizer containers provide buffer volumes for the thermal expansion of Gd-LS. However, the buffer volumes could be overwhelmed for the case that the inverse siphon is broken, so that  the temperature inside the detector is monitored by eight PT100 RTD sensors installed in the veto region. Four RTDs are installed in the detector bottom region, and another four RTDs in the middle barrel region. An NI 9216 module reads measured temperatures out of RTDs. Figure~\ref{fig:position_sensor} shows their installed sensor locations.

\subsection{Detector pressure monitor}
The JSNS$^{2}$ detector is hermetically sealed to reduce the (Gd-)LS’s oxygen exposure. The air tightening allows for pressure differences to develop between the detector and the surrounding atmosphere. The stainless steel tank is strong enough to withstand pressure differences up to 20 kPa \cite{cite:JSNS2_Veto_detector}. Two types of relative pressure meters of GC-31 \cite{cite:GC_31} and GC-62 \cite{cite:GC_62} are installed to monitor the pressure difference. The effective ranges are 100 kPa for GC-31 and 2 kPa for GC-62. The GC-62 sensor monitors the air pressure difference between the target chimney and the veto region. The GC-31 sensor measures the air pressure difference between the detector inside and outside. The pressure sensors provide a voltage output of 1 to 5V proportional to the measured pressure difference. The output voltages are read out by the NI 9201 module.

\subsection{Ambient sensor}
A TR-73U sensor \cite{cite:TR-73U} is used to monitor environmental conditions around the detector. The sensor measures the temperature, humidity, and atmospheric pressure in the experimental area near the detector. The obtained data are read out by the LabVIEW program via an RS232 connector adapted to USB. The measured results are also displayed on a LCD panel.

\section{Visualization of monitoring data}
\label{sec:Grafana}
\begin{figure}[h]
\begin{center}
\includegraphics[scale=0.4]{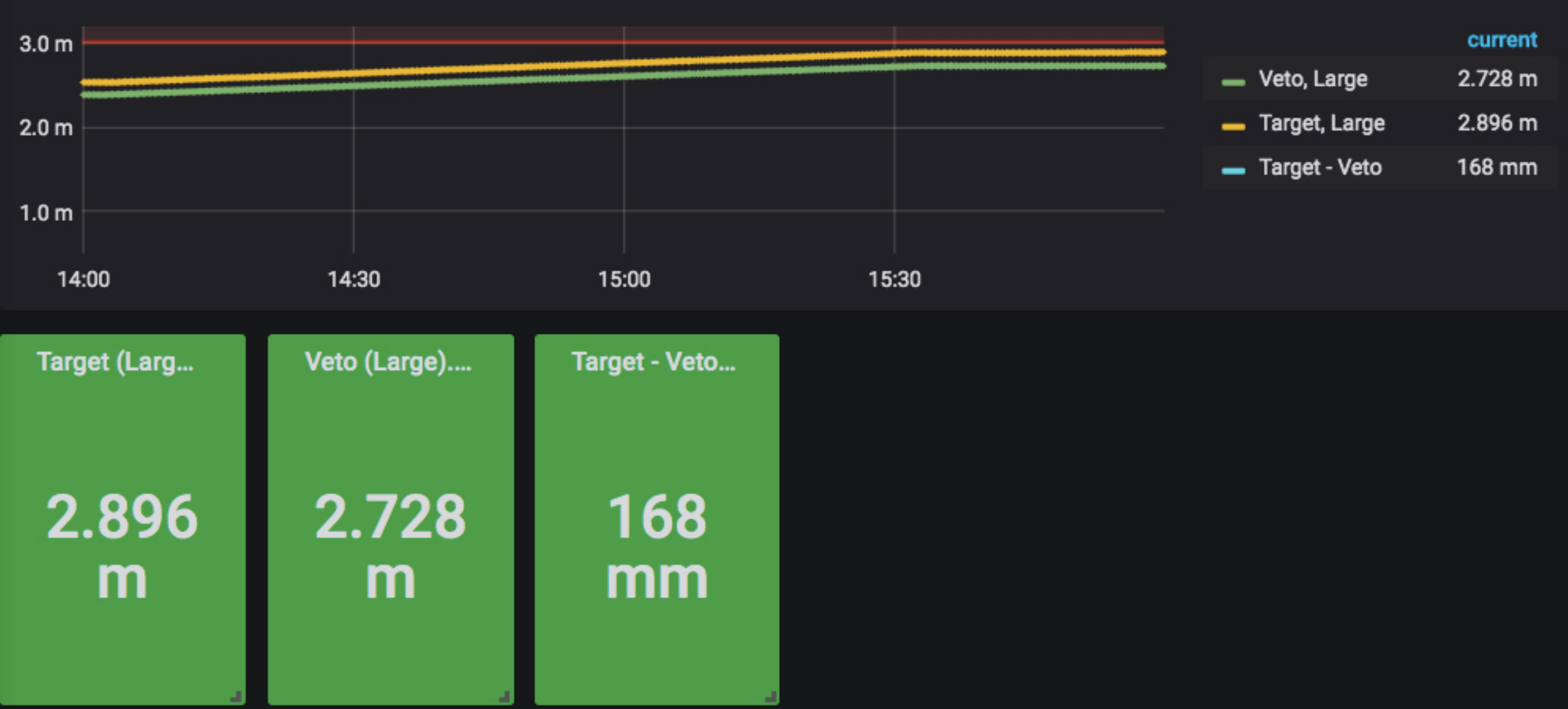}
\end{center}
\caption{\setlength{\baselineskip}{4mm} Screenshot of SCMS display by Grafana during liquid filling. The top panel shows a chronological graph of measured liquid levels and their difference inside the detector, and bottom colored panels show the latest results for comparison.}
\label{fig:Grafana_SCM}
\end{figure}

\begin{figure}[h]
\begin{center}
\includegraphics[scale=0.4]{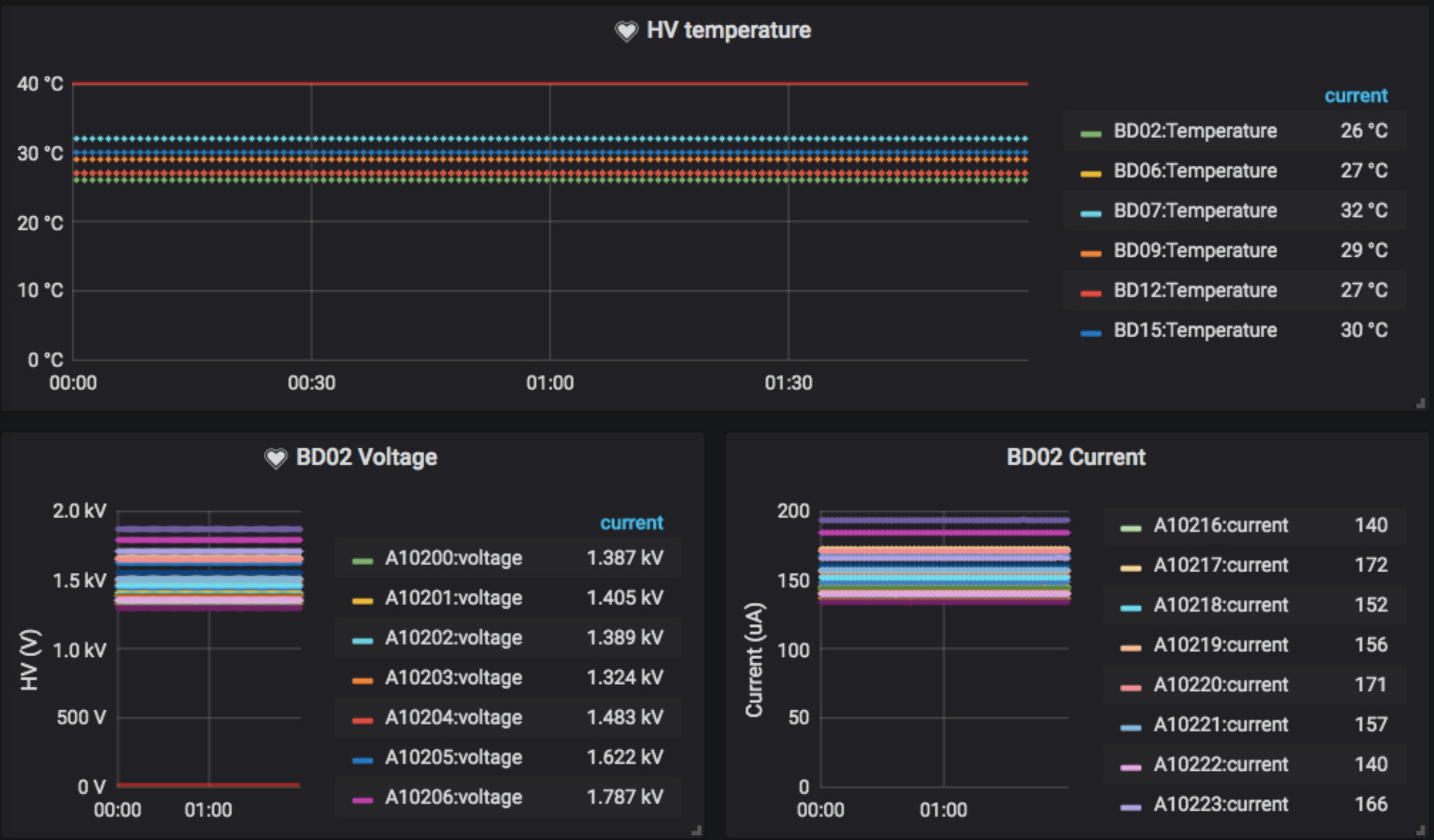}
\end{center}
\caption{\setlength{\baselineskip}{4mm} Screenshot of HV control and monitoring display by Grafana. The top panel shows the temperature of each HV supply module. The bottom left panel shows the supplied HV of 24 channels, and the bottom right shows the electric currents of the 24 channels.}
\label{fig:Grafana_HVCM}
\end{figure}

A Grafana graphical user interface is used to display the data recorded in the MySQL database and issues an alarm if necessary. Figure~\ref{fig:Grafana_SCM} shows a screenshot of the SCMS display during liquid filling. It displays measured liquid levels and their difference. If the difference exceeds a preset threshold, the panel color changes to red and the Grafana sends warning e-mails and SNS messages. Figure~\ref{fig:Grafana_HVCM} shows a screenshot of the monitored HV values and temperatures of HV supply modules. If the measured HV value of any channel becomes zero, Grafana displays an alarm signal and sends warning emails and SNS messages.

\section{Summary}
The JSNS$^2$  detector was complete and is operated for data-taking in search for sterile neutrino oscillation at J-PARC. For reliable control and quick monitoring of the detector operational status, we have successfully installed various sensors with appropriate readout modules and a LabVIEW based monitoring display using a Grafana GUI. The sensor readout data are recorded into a MySQL database. The SCMS also issues alarms to alert users if any of the monitored values are out of their preset range. The first JSNS$^2$ run was completed in June 2020 with successful operation of the SCMS. It demonstrates a reliable and robust SCMS performance for the JSNS$^2$ experiment.

\section*{Acknowledgements}
We thank the J-PARC staff for their support. We acknowledge the support of the Ministry of Education, Culture, Sports, Science, and Technology (MEXT) and the JSPS grants-in-aid (Grant Number 16H06344, 16H03967, 20H05624), Japan. This work is also supported by the National Research Foundation of Korea (NRF) Grant No. 2016R1A5A1004684, 2017K1A3A7A09015973, 2017K1A3A7A09016426, 2019R1A2C3004955, 2016R1D1A3B02010606, 2017R1A2B4011200, 2018R1D1A1B07050425, 2020K1A3A7A0908
0133 and 2020K1A3A7A09080114. Our work has also been supported by a fund from the BK21 of the NRF. The University of Michigan gratefully acknowledges the support of the Heising-Simons Foundation. This work conducted at Brookhaven National Laboratory was supported by the U.S. Department of Energy under Contract DE-AC02-98CH10886. The work of the University of Sussex is supported by the Royal Society grant no.IES\textbackslash R3\textbackslash 170385.

\end{document}